\documentclass{article}
\usepackage{epsfig}
\usepackage[english]{babel}
\usepackage{color}
\newcommand{\parl}{\parallel}
\usepackage{graphicx}
\usepackage{amsmath}
\usepackage{subcaption}
\begin{document}
\title{Excitonic Spectra of Wide Parabolic Quantum Wells}
\author{Gerard Czajkowski, Sylwia Zieli\'{n}ska-Raczy\'{n}ska, and David
Ziemkiewicz\footnote{david.ziemkiewicz@utp.edu.pl} \\
\\
\emph{Institute of Mathematics and Physics, UTP University of Science and Technology},\\\emph{ Al. Prof. S. Kaliskiego 7, 85-789 Bydgoszcz, \emph{Poland}}} 
\date{\vspace{-5ex}}
\maketitle
\begin {abstract}
The optical properties of wide Quantum Wells are considered,
taking into account the screened electron-hole interaction
potential and parabolic confinement potentials, different for the
electrons and for the holes. The role of the interaction
potential which mixes the energy states according to different
quantum numbers is stressed. The results obtained by our method are in agreement with the observed spectra
and give the possibility to the assessment of the resonances.
\end {abstract}
\section{Introduction}
{A single Quantum Well (QW) is formed when a thin layer of a
narrow gap semiconductor material (Well) lies between layers of
wider gap materials (Barriers). The size of QWs (taken in the
growth direction) is generally between 1 and 20 nm (narrow QWs).
Mostly the size is comparable with the effective excitonic Bohr
 radius $a^*$ of the QW material (for example, $a^*$ is about 10 nm
for GaAs). When the size of the QW is of the order of a few
excitonic Bohr radii, we speak about Wide Quantum Well (50-100 nm
in the case of GaAs Wells). Typical property of the QWs is the
confinement of the carriers (electrons and holes) inside the well.
Mostly one considers a rectangular shape of the confinement
potential, but also other shapes are considered. A special
fabrication technique allows to create a parabolic shaped
confinement, leading to the so-called Parabolic Quantum Wells
(PQWs). More generally, we can speak about Parabolic Quantum
Nanostructures (for example, Parabolic Quantum Dots). When the
size of a PQW corresponds to the Wide QW region, we deal with Wide
Parabolic Quantum Wells (WPQWs). } The structures with parabolic
confinement have attracted more attention in the recent decades
 (for example, Ref.
\cite{MillerGossard}-\cite{Taqi}).

Each type of semiconductor structures has a specific property
depending on its dimensionality, which influences the optical
characteristic of a given structure.
 %Considering the optical properties of semiconductor structures
%with different dimensionality, beginning with bulk crystals and
%going over to low dimensional systems we state that every type of
%structure has a specific property which is dominant in determining
%the optical properties.
 In bulk crystals the coupling between the
external electromagnetic wave and the internal polariton modes
gives rise to the so-called ABC problem, which was extensively
examined in the past \cite{Birman82}-\cite{Agran2009}). In
low-dimensional structures as Quantum Wells, Quantum Wires and
Quantum Dots mostly the long-wave approximation is used, so the
electrodynamical aspect is simplified, however, on the other hand,
due to the confinement, the separation of center-of-mass and
relative e-h motion is not possible and the problem of solving
6-dimensional Schr\"{o}dinger equation appears. The additional
difficulty consists on the fact that the spherical symmetry of the
Coulomb potential is not compatible with the cylindrical symmetry
of most nanostructures. Thus several approximations have been
proposed which used the lowering of the dimension of the problem
(for example, \cite{CzSil2006}). Recently, also due to the
increasing power of computers, direct numerical solution of the
6-dimensional 2-particles Schr\"{o}dinger equation has been
performed (for example, Refs. \cite{Tadic_rings},
\cite{Schillak_epj}).

Here we consider Wide Quantum Wells (WQWs), nanostructures where
the optical active layers have extension of a few excitonic Bohr
radii in the growth direction.  To a certain approximation, the
polaritonic aspect can be here neglected, and the long-wave
approximation sustained. The exciton is not squeezed as in other
low-dimensional structures. In typical Quantum Wells with the
dimension of, say, one excitonic Bohr radius in the growth
direction we observe only a few excited states. In WQWs, due to
the greater extension, significantly larger number of states is
observed. The Coulomb potential and different confinement
potentials for electrons and holes couples electron and hole
confinement states of different quantum numbers. Such phenomena
have been observed experimentally (see, for example, Ref.
\cite{MillerGossard}). We propose the computational method which
leads to analytical expression for the electric susceptibility of
wide parabolic quantum well taking into account the screened
electron-hole interaction and parabolic confinement potential.
With the purpose of exemplification, we consider a quantum well
with GaAs as the optically active layer and Ga$_{1-x}$Al$_x$As as
the barriers, where the active layer is of the extension of a few
excitonic Bohr radii. The absorption spectra of such a structure
show a large number of resonances ($n=8$ observed in
\cite{MillerGossard}). The choice of optimal effective potential
parameters as well as the damping constant used in our calculation
is verified by numerical calculations of the total fitting error
for maxima of susceptibility. We have chosen as reference the
paper by Miller \emph{et al} \cite{MillerGossard} because it
contains a lot of experimental data which allowed to compare the
obtained theoretical results with experiment. The agreement
between our calculated spectrum and experimental data is very good
with regard to the number and position of the maxima of
susceptibility.

Our paper is organized as follows. In the section 2,
 we present the assumptions of considered model and solve the
 constitutive equation with effective electron-hole interaction potential.
 Section 3 is devoted to the details of the applied potential. Next, in section 4,
  the derived solution of constitutive equation is used to obtain the energy levels of the considered
  GaAs/Ga$_{1-x}$Al$_x$As
  wide parabolic quantum well. Finally, in section 5, the susceptibility for such nanostructure is calculated and discussed.
   The comparison of obtained results with experimental data and a brief overview of optimizing procedure is included.

\section{The Model}
We will compute the linear optical response of a WPQW to  a plain
electromagnetic wave
\begin{equation}\label{stpar_wave}
E_i(z,t)=E_{i0}\exp({\rm i}k_0z-{\rm i}\omega t), \qquad
k_0=\frac{\omega}{c},
\end{equation}
\noindent attaining the boundary surface of the WPQQW active layer
located at the plane  $z=0$. The second boundary is located at the
plane
 $z=L$. In the case of GaAs Well the extension $L$ will be of the order
2-40
 nm.

Due to this extension, the following aspects should be taken into
account. 1) Several confinement states resulting from the
confinement in the $z$ direction are to be included in the
consideration. 2) The electron-hole potential plays an important
role and cannot be approximated by a 2-dimensional potential, as
was sometimes done in the case of simple Quantum Wells (the limit
$L\rightarrow 0$). 3) The parabolic shape of the confinement
potential is assumed
\begin{equation}\label{stpar_potencjaluwiezienia}
V_{\rm
uw}(z_e,z_h)=\frac{1}{2}m_{ez}\omega_e^2z_e^2+\frac{1}{2}m_{hz}\omega_h^2z_h^2.
\end{equation}
4) The electron-hole interaction is described by the potential
$V({\bf r}_e,{\bf r}_h)$. 5) We adopt the real density matrix
approach to compute the optical properties. In this approach the
linear optical response will be described by a set of coupled
equations: two constitutive equations for the coherent amplitudes
$Y_\nu(\textbf{r}_e,\textbf{r}_h)$, $\nu=\hbox{H,L}$ stands for
heavy-hole (H) and light-hole exciton); from them the polarization
can be obtained and used in Maxwell's field equations.  Having the
field we can determine the QW optical functions (reflectivity,
transmission, and absorption).

Thus the next steps are the following: We formulate the
constitutive equations. The equations will be then solved giving
the coherent amplitudes $Y$. From the amplitudes we compute the
polarization inside the Quantum Well, the electric field of the
wave, and the optical functions. This scheme will be applied for
the case investigated in Ref.~\cite{MillerGossard}.

%\section{The constitutive equations for a Wide Parabolic Quantum %Well}\label{Constitutive.equation}

As was explained in, for example, Ref. \cite{RivistaGC}, the
constitutive equation for the coherent amplitude
 $Y$ in a Quantum Well has the form
\begin{equation}\label{stpar_konstytutpow}
\left[E_{g}-\hbar\omega-{\rm
i}{\mit\Gamma}+\frac{\hat{p}_{ez}^2}{2m_{e}}+\frac{\hat{p}_{hz}^2}{2m_{hz}}+\frac{\hat{\bf
p}_{\rho}^2}{2\mu_{\parl }}+\frac{\hat{\bf
p}_{\parl}^2}{2M_{\parl }}+V_{eh}(\rho,z_e,z_h)+
V_{\rm conf}(z_e,z_h)\right]Y
= {\bf M}({\bf r}){\bf E}({\bf R}),
\end{equation}
\noindent where ${\bf M}({\bf r})$ is the transition dipole
density, which form we have assumed as
\begin{eqnarray}\label{dipoledensity}
&&{\bf M}({\bf r})={\bf M}(\rho,z,\phi)
=\frac{\textbf{M}_{0}}{2\pi\rho_{0}}\delta(z)\delta\left(\rho-\rho_{0}\right),
\end{eqnarray}
\noindent  $z=z_e-z_h$ being the relative coordinate in the $z$
direction, $\rho_{0}$ is the coherence radius, ${\bf R}$ jest is
the excitonic center-of-mass coordinate and ${\bf E}({\bf R})$ is
the electric field vector of the wave propagating in the QW;
$V_{\rm conf}(z_e,z_h)$ is the confinement potential for electrons
and holes, and  $\hat{\bf p}_{\rho}$, $\hat{\bf p}_{\parl}$  are
the momentum operators for the excitonic relative- and
center-of-mass motion in the QW plane.

In the following we assume that the
propagating wave is linearly polarized in the $x$ direction, and
that the vector M has a non-vanishing component in the same
direction. Taking the confinement potential in the form
(\ref{stpar_potencjaluwiezienia}) we find in the equation
(\ref{stpar_konstytutpow}) Hamilton operators for the
one-dimensional harmonic oscillator
\begin{equation}
H_e=\frac{\hat{p}_{ez}^2}{2m_{e}}+\frac{1}{2}m_e\omega_e^2z_e^2,\,\,\,\,\,\,
H_h=\frac{\hat{p}_{hz}^2}{2m_{h}}+\frac{1}{2}m_h\omega_h^2z_h^2.
\end{equation}
\noindent Therefore we look for a solutions $Y$ in terms of the
eigenfunctions of the operators $H_e,H_h$
\begin{equation}\label{stpar_rozwiniecie1}
Y\left(\rho,z_e,z_h\right)=\sum\limits_{j,n=0}^{N}\psi_{ej}(z_e)\psi_{nh}(z_h)Y_{jn}(\hbox{\boldmath
$\rho$}),
\end{equation}
\noindent The eigenfunctions $\psi_j$ have the form
\begin{eqnarray}\label{stpar_eigenf1doscillator8}
 \psi_{ej}(z_e)&=&
 \pi^{-1/4}\sqrt{\frac{\alpha_{e}}{2^j j!}} H_j\left(\alpha_{e}z_e\right)
e^{-\frac{\alpha_{e}^2}{2}z_e^2};\qquad
 \alpha_{e} = \sqrt{\frac{m_{ez} \omega_{e}}{\hbar}},\nonumber\\
\psi_{nh}(z_h)&=&
 \pi^{-1/4}\sqrt{\frac{\alpha_{h}}{2^n n!}} H_n\left(\alpha_{h}z_h\right)
e^{-\frac{\alpha_{h}^2}{2}z_h^2};\qquad
 \alpha_{h} = \sqrt{\frac{m_{hz} \omega_{h}}{\hbar}},
\end{eqnarray}

\noindent with the Hermite polynomials $H_n(x)$, and the
corresponding eigenvalues $E_{n}=\left(n+\frac{1}{2}\right)\hbar\omega$.
\noindent Substituting (\ref{stpar_rozwiniecie1}) into the eq.
 (\ref{stpar_konstytutpow}) we obtain equations for the functions
 $Y_{jn}$
\begin{eqnarray}\label{stpar_konstytutpow1}
\sum\limits_{j,n=0}^{N}\left[E_{g}-\hbar\omega-{\rm
i}{\mit\Gamma}+E_{je}+E_{nh}+\frac{\hat{\bf p}_{\hbox{\boldmath
$\rho$}}^2}{2\mu_{\parl }}+\frac{\hat{\bf
p}_{\parl}^2}{2M_{\parl }}+V_{eh}(\hbox{\boldmath $\rho$},z_e,z_h)
\right]\psi_j(z_e)\psi_{n}(z_h)Y_{jn}(\hbox{\boldmath
$\rho$})
= {\bf M}({\bf r}){\bf E}({\bf R}).\nonumber\\
\end{eqnarray}
Now we have to specify the shape of the interaction potential
$V_{eh}(\hbox{\boldmath $\rho$},z_e,z_h)$ and the wave electric
field ${\bf E}({\bf R})$. We assume the so-called long-wave
approximation and consider ${\bf E}({\bf R})$ in the equation
 (\ref{stpar_konstytutpow1}) as a constant quantity.The electron-hole interaction potential $V_{eh}(\hbox{\boldmath
$\rho$},z_e,z_h)$ is, in general, the screened Coulomb potential

\begin{equation}\label{stpar_potencjalcoulomba}
V_{eh}(\rho,z_e,z_h)=-\frac{e^2}{4\pi\epsilon_b\sqrt{\rho^2+\left(z_e-z_h\right)^2}},
\end{equation}
\noindent $\epsilon_b$ being the dielectric constant of the QW
material. Despite of the nanostructures with
cylindrical symmetry considered in ref.\cite{Schillak_epj}, in the
case of the wide QWs one does not have an orthonormal basis of
functions so the use of an effective e-h interaction potential
will be made
\begin{equation}\label{stpar_potencjalharmoniczny}
V_{eh}=-S\exp\left[-v\left(z_e-z_h\right)^2-w\rho^2\right].
\end{equation}
\noindent where  $v,w$ are certain parameters which will be
estimated below. Using the above potential, the dipole density
(\ref{dipoledensity}), and neglecting the center-of-mass in plane
motion, we put the constitutive equation
(\ref{stpar_konstytutpow}) into the form
\begin{equation}\label{stpar_konstytuwn3}
\left(E_{rs}+\frac{\hat{\bf p}_{\hbox{\boldmath
$\rho$}}^2}{2\mu_{\parl
}}\right)Y_{rs}-e^{-w\rho^2}\sum\limits_{nj}V_{rsnj}Y_{nj}=E\frac{M_0}{2\pi\rho_0}\langle
r\vert s\rangle\delta\left(\rho-\rho_0\right),
\end{equation}
where
\begin{eqnarray}\label{stpar_parzystosc}
E_{rs}&=&E_g+E_{re}+E_{sh}-\hbar\omega-{\rm
i}{\mit\Gamma},\qquad r,s,=0,1,2,\ldots\,,\nonumber\\
\label{stpar_elementymacierzowe}
V_{rsnj}&=&S\langle rs\left|\exp\left[-v\left(z_e-z_h\right)^2\right]\right\vert nj\rangle\\
\end{eqnarray}
\noindent With regard to the shape of the functions  $\psi$ only
states of the same parity will give nonvanishing elements $\langle
r\vert s\rangle$ so  the states $\vert 0e0h\rangle, \vert
0e2h\rangle, \vert 1e3h\rangle$ etc. will be taken into account.
To summarize in order to calculate the optical response of a wide
Quantum Well it is necessery to solve the constitutive equation
(\ref{stpar_konstytuwn3}) using the matrix elements $\langle
r\vert s\rangle$ and the potential matrix elements
(\ref{stpar_elementymacierzowe}).

\section{The parameters of the effective potential}
The further calculations require the estimation of parameters
characterizing the effective potential
(\ref{stpar_potencjalharmoniczny}). We make the following
assumptions: 1) The potential is isotropic, in analogy to the
Coulomb potential in isotropic materials. The nanostructure
anisotropy is included in the quasiparticles effective masses.
This assumption leads to the equality $u=v$. 2) We assume the
value $S \approx 2R^*$ ($R^*$ being the effective excitonic
Rydberg energy for the given crystal); the exact value $S$ will be
established later.
 We determine the ground state energy of a hydrogen-like atom,
 where the interaction between the charges is given by
(\ref{stpar_potencjalharmoniczny}). To this end we solve the
Schr\"{o}dinger equation
\begin{equation}\label{stpar_schroedwodor}
-\frac{\hbar^2}{2\mu}\left(\frac{{\rm d}^2}{{\rm
d}r^2}+\frac{2}{r}\frac{\partial}{\partial
r}\right)\psi-SR^*e^{-vr^2}\psi=E\psi.
\end{equation}
\noindent Making use of the relation $
\frac{\hbar^2}{2\mu}=R^*a^{*2}$
 with the effective Bohr radius  $a^*$, we introduce
scaled variables
\begin{equation}\label{stpar_zmienneskalowane}
\rho=\frac{r}{a^*},\qquad\varpi=wa^{*2},\qquad\varepsilon=\frac{E}{R^*},
\end{equation}
\noindent transforming the eq.(\ref{stpar_schroedwodor}) to
$H\psi=\varepsilon\psi$ with the Hamiltonian
\begin{equation}\label{stpar_hamiltschroedwodorskal}
H=-\left(\frac{{\rm d}^2}{{\rm d}\rho^2}+\frac{2}{\rho}\frac{\rm
d}{{\rm d}\rho}\right)-Se^{-\varpi\rho^2}.
\end{equation}
\noindent The considered Schr\"{o}dinger equation will be solved
by the variational method. Using the trial function
\begin{equation}\label{stpar_psiproba}
\psi=e^{-\lambda\rho^2/2},
\end{equation}
we arrive at
\begin{equation}\label{stpar_varepslambda1}
\varepsilon(\lambda)=\frac{3}{2}\lambda-S\left(\frac{\lambda}{\lambda+\varpi}\right)^{3/2}.
\end{equation}
By assuming the condition $\epsilon=-1$ and the vanishing derivative $\epsilon'=0$, for any given value of $\lambda,$ one obtains a system of equations for two unknown quantities $S$ and $\varpi$:
\begin{eqnarray}\label{stpar_ukladrownanlamda}
\frac{3}{2}\lambda-S\left(\frac{\lambda}{\lambda+\varpi}\right)^{3/2}&=&-1,\nonumber\\
\frac{1}{S}-\frac{\varpi}{\lambda^2}\left(1+\frac{\varpi}{\lambda}\right)^{-5/2}&=&0,
\end{eqnarray}
\noindent and their values  will be than used to determine the
elements (\ref{stpar_elementymacierzowe}). Looking for  a solution
which will reproduce the exact energy value $\varepsilon=-1$ we
choose $\lambda=0.34$, $S =2.22$, and $\varpi=0.1$.
In order to  compute the optical spectra  we have to solve the
system  (\ref{stpar_konstytuwn3}) of coupled differential
equations, but it will be easier to obtain the solutions by
transforming the equations into linear algebraic equations. This
can be done in the following way. Assume, for a moment, that the
equation with indices (0,0) decouples from the remaining
equations. Denoting $V_{0000}=V_0$ we obtain the following
equation for the amplitude $Y_{00}$
\begin{eqnarray}
\left[E_g+E_{0e}+E_{0h}-\hbar\omega-{\rm i}{\mit\Gamma}-\frac{\hbar^2}{2\mu_\parl}\left(\frac{{\rm d}^2}{{\rm d}\rho^2}+\frac{1}{\rho}\frac{\rm d}{{\rm d}\rho}+\frac{1}{\rho^2}\frac{{\rm d}^2}{{\rm d}\phi^2}\right)-V_0e^{-w\rho^2}\right]Y_{00}
=M_0\frac{\delta\left(\rho-\rho_0\right)}{2\pi\rho_0}E\langle
0\vert 0\rangle.
\end{eqnarray}
\noindent After  rescaling the spatial
variables in the effective excitonic Bohr radius the above
equation becomes

\begin{equation}\label{stpar_rownaniey_00}
k_{00}^2Y_{00}+\left(-\frac{{\rm d}^2}{{\rm
d}\rho^2}-\frac{1}{\rho}\frac{\rm d}{{\rm
d}\rho}-\frac{1}{\rho^2}\frac{{\rm d}^2}{{\rm
d}\phi^2}-v_0e^{-\varpi
\rho^2}\right)Y_{00}=\frac{2\mu_\parl}{\hbar^2}M_0E\frac{\delta\left(\rho-\rho_0\right)}{2\pi\rho_0}\langle
0\hbox{e}\vert 0\hbox{h}\rangle,
\end{equation}
\noindent where now $\rho$ denotes the scaled variable $\rho/a^*$,
and
\begin{equation}
k_{00}^2=\frac{E_g+E_{0e}+E_{0h}-\hbar\omega-{\rm
i}{\mit\Gamma}}{R^*},\qquad v_0=\frac{V_0}{R^*}.
\end{equation}
Assuming the $s$-symmetry for the ground state, we first solve the
Schr\"{o}dinger equation

\begin{equation}\label{stpar_schroed00}
\left(-\frac{{\rm d}^2}{{\rm d}\rho^2}-\frac{1}{\rho}\frac{\rm
d}{{\rm d}\rho}-v_0e^{-\varpi \rho^2}\right)\psi=\varepsilon\psi.
\end{equation}
\noindent Using  the variational method we  solve above equation,
using the trial function
%\begin{equation}\label{trial}
$\psi=e^{-\lambda\rho^2/2}$.
%\end{equation}
\noindent{Denoting by $H$ the left-hand-side operator, we compute
the expression which should be minimized
\begin{equation}\label{stpar_varepsilonzero}
\varepsilon(\lambda)=\frac{\langle \psi \vert
H\psi\rangle}{\langle \psi\vert\psi\rangle}=\lambda-\frac{\lambda
v_0}{(\lambda+\varpi)}.
\end{equation}}
\noindent The condition for the minimum yields
\begin{equation}\label{stpar_kwadratowe00}
(\varpi-v_0)x^2+2\varpi x+\varpi=0,
\end{equation} where $x=\frac{\varpi}{\lambda}.$
The function
\begin{equation}
\psi_0(\rho,\phi)=\frac{\sqrt{2\lambda}}{\sqrt{2\pi}}e^{-\lambda\rho^2/2},
\end{equation}
\noindent with the value of $\lambda$ obtained from the above equation can be
considered as the normalized eigenfunction of the Schr\"{o}dinger
equation with the Hamiltonian
\begin{equation}\label{stpar_hamiltonianzero}
H_0=-\frac{{\rm d}^2}{{\rm d}\rho^2}-\frac{1}{\rho}\frac{\rm
d}{{\rm d}\rho}-v_0e^{-\varpi \rho^2}=\hat{\bf
p}_\rho^2-v_{0000}e^{-\varpi \rho^2}.
\end{equation}
\noindent The index 0 denotes that this is the lowest energy state
for the relative electron-hole motion with the assumed effective
e-h interaction potential. Now we put $Y_{00}(\rho)$ into the form
\begin{equation}
Y_{00}(\rho)=A\psi_0(\rho).
\end{equation}
With regard to
%\begin{equation}\label{stpar_schroedingerzero}
$H_0\psi_0=\varepsilon_0\psi_0$,
%\end{equation}
\noindent where $\epsilon_0=\epsilon$ corresponds to the above
estimated energy value, we obtain from (\ref{stpar_rownaniey_00})
\begin{equation}
A=\frac{1}{k_{00}^2+\varepsilon_0}\frac{2\mu_\parl}{\hbar^2}M_0E\langle
0\vert 0\rangle \psi_0(\rho_0),
\end{equation}
\noindent and  the amplitude $Y(\rho, z_e,z_h)$ has the form
\begin{equation}
Y(\rho,
z_e,z_h)=\frac{1}{k_{00}^2+\varepsilon_0}\frac{2\mu_\parl}{\hbar^2}M_0E\langle
0\hbox{e}\vert 0\hbox{h}\rangle
\psi_0\left(\rho_0\right)\psi_0(\rho)\psi_{0e}\left(z_e\right)\psi_{0h}\left(z_h\right).
\end{equation}
\section{The solution of the constitutive equation}
Making use of the above calculated function  $\psi_0$, we put the
amplitude (\ref{stpar_rozwiniecie1}) into the form
\begin{equation}\label{stpar_rozwiniecie2}
Y\left(\rho,z_e,z_h\right)=\psi_0(\rho)\sum\limits_{j,n=0}^{N}\psi_{ej}(z_e)\psi_{nh}(z_h)Y_{jn},
\end{equation}
where now $Y_{jn}$ are constant coefficients. Equation
(\ref{stpar_konstytuwn3}) takes now the form

\begin{equation}\label{stpar_konstytuwn4}
\left(E_{rs}+\frac{\hat{\bf p}_{\hbox{\boldmath
$\rho$}}^2}{2\mu_{\parl
}}\right)\psi_0(\rho)Y_{rs}-e^{-\varpi\rho^2}\psi_0(\rho)\sum\limits_{nj}V_{rsnj}Y_{nj}=E\frac{M_0}{2\pi\rho_0}\langle
\hbox{e}r\vert\hbox{h}s\rangle\delta\left(\rho-\rho_0\right).
\end{equation}
\noindent After rescaling the spatial variable $\rho\rightarrow
\rho/a^*$  we obtain from (\ref{stpar_konstytuwn4}) the relation
\begin{equation}\label{stpar_konstytuwn5}
\left(k_{rs}^2+\hat{\bf
p}_\rho^2\right)\psi_0(\rho)Y_{rs}-e^{-\varpi\rho^2}\psi_0(\rho)\sum\limits_{nj}v_{rsnj}Y_{nj}=\frac{2\mu_\parl}{\hbar^2}E\frac{M_0}{2\pi\rho_0}\langle
\hbox{e}r\vert\hbox{s}\rangle\delta\left(\rho-\rho_0\right),
\end{equation}
which, using the quantities
$k_{rs}^2=\frac{E_{rs}}{R^*},v_{rsnj}=\frac{V_{rsnj}}{R^*}$ can be
written as
\begin{eqnarray}\label{stpar_konstytuwn6}
&&\left(k_{rs}^2+\hat{\bf p}_\rho^2-v_{0000}e^{-\varpi\rho^2}\right)\psi_0(\rho)Y_{rs}+v_{0000}e^{-\varpi\rho^2}\psi_0(\rho)Y_{rs}\nonumber\\
&&-e^{-\varpi\rho^2}\psi_0(\rho)\sum_{nj}v_{rsnj}Y_{nj}=\frac{2\mu_\parl}{\hbar^2}E\frac{M_0}{2\pi\rho_0}\langle
\hbox{e}r\vert\hbox{h}s\rangle\delta\left(\rho-\rho_0\right),
\end{eqnarray}
\noindent and, in consequence,
\begin{equation}\label{stpar_konstytuwnuklad1}
\left(k_{rs}^2+\epsilon_0\right)Y_{rs}+v_{0000}\frac{\lambda}{\lambda+\varpi}Y_{rs}
-\frac{\lambda}{\lambda+\varpi}\sum\limits_{nj}v_{rsnj}Y_{nj}=\frac{2\mu_\parl}{\hbar^2}E{M_0}\langle
\hbox{e}r\vert\hbox{h}s\rangle\psi_0\left(\rho_0\right).
\end{equation}
\noindent We obtained a system of linear algebraic equations for
the coefficients $Y_{nj}$. Having them, we determine the amplitude
 $Y$ (or amplitudes, when accounting the heavy- and light hole excitons H and L. Given the amplitude, we compute the polarization inside the quantum well
 and the electric field. For the further calculations we introduce dimensionless
 quantities $\mathcal{Y}_{rs}$
\begin{equation}
\frac{2M_0}{\epsilon_0\epsilon_b\pi a^*}Y_{rs} =
\mathcal{Y}_{rs}\cdot E
\end{equation}

%\noindent Multiplying both sides of eq. by (\ref{stpar_konstytuwnuklad1})
%\begin{equation}
%\frac{2M_0}{\epsilon_0\epsilon_b\pi a^*}
%\end{equation}

\noindent and arrived to the formula %we obtain
\begin{equation}\label{stpar_niemianowane}
\left(k_{rs}^2 + \epsilon_0\right)\mathcal{Y}_{rs} +
v_{0000}\frac{\lambda}{\lambda + \varpi}\mathcal{Y}_{rs} -
\frac{\lambda}{\lambda + \varpi}\sum_{nj}v_{rsnj}\mathcal{Y}_{nj}
= \frac{\Delta_{LT}}{R^*}\langle
\hbox{e}r\vert\hbox{h}s\rangle\psi_0\left(\rho_0\right),
\end{equation}
where we used the relation
%\begin{equation}
$2\frac{2\mu_\parl}{\hbar^2}\frac{M_0^2}{\epsilon_0\epsilon_b\pi
a*} = \frac{\Delta_{LT}}{R*},$
%\end{equation}
with $\Delta_{LT}$ being the transversal-longitudinal splitting energy, (see for example,\cite{RivistaGC}).
%\section{The parameters of the confinement potential}
The described method can be used when we define the confinement
energies $\hbar\omega_e, \hbar\omega_h$ and thus the parameters
$\alpha_e,\alpha_h$ We will choose them to compare our theoretical
results with the experimental findings of Miller et
al.\cite{MillerGossard}. They obtained optical spectra for
GaAs(Well)/Ga$_{0.7}$Al$_{0.3}$As (Barrier) QWs of three
thicknesses: $L=51 \pm~3,5~\hbox{nm}, L=32,5 \pm~3.5~\hbox{nm},
L=33.6 \pm~3.5~\hbox{nm}$. It can be noticed the uncertainty in
determining the well thickness. The confinement parameters are
obtained as follows. We consider a symmetric QW with a rectangular
confinement potential $V$
\begin{equation}
V = E_g(\hbox{Ga}_{0,7}\hbox{Al}_{0,3}\hbox{As}) -
E_g(\hbox{GaAs}) = 482.8~\hbox{meV},
\end{equation}
see Table \ref{parametervalues}. The confinement potentials for
electrons $V_e$ and holes $V_h$ are chosen as
\begin{equation}\label{stpar_podzial_potencjalu}
V_e = 0.85\,V = 410.38~\hbox{meV},\qquad V_h =0.15\,V=
72.42~\hbox{meV}.
\end{equation} Then we compute the lowest energy states in the QW with potentials $V_e, V_h$.
We follow the scheme from Ref. \cite{Davies} where the lowest
energies result from the equation

\begin{equation}\label{stpar_stud_prost}
\left\{\left[\frac{m_W}{m_B}\left(\frac{\theta_0^2}{\theta^2}-1\right)\right]+1\right\}^{-1}
- \cos^2\theta = 0,
\end{equation}
where the dimensionless parameters $\theta, \theta_0$ are defined
as
\begin{equation}\label{deftheta}
\theta = \frac{kL}{2} = \frac{1}{2}\sqrt{\frac{E}{R^*}}\frac{L}{a^*},\,\,\,\,\,\,
\theta_0^2 = \frac{m_WVL^2}{2\hbar^2} =
\frac{1}{4}\left(\frac{V}{R^*}\right)\left(\frac{L}{a^*}\right)^2,
\end{equation}
and the index $W$ means Well; the values $a^*$, $R^*$ are
appropriate for electrons and holes for the QW material, and are
defined as
\begin{equation}\label{rydberg}
 R^*=\frac{me^4}{2(4\pi\epsilon_0\epsilon_b)^2\hbar^2},\,\,\,\,\,\,\,\,
 a^*=\frac{\hbar^2(4\pi\epsilon_0\epsilon_b)}{m e^2}.
 \end{equation}
 The below listed values are obtained when we insert in
 (\ref{rydberg}) the appropriate effective masses: $m_e$ for
 $R^*_e,a^*_e$, and $\mu_{\parallel H,L}$ for $R^*_H, a^*_H$ and
 $R^*_L, a^*_L$; $\mu_{\parallel H,L}$ are the in-plane reduced masses
 for the electron-hole pair and for the heavy- and light-hole
 exciton data.
\begin{table}[h]
\caption{Band parameter values for GaAs, AlAs, and
Ga$_{0,7}$Al$_{0,3}$As,
 AlAs data from~\protect\cite{GrundmannStier}, for Ga$_{0.7}$Al$_{0.3}$As
by linear interpolation. Energies in meV, masses in free electron
mass $m_0$, $\gamma_1, \gamma_2$ are Luttinger parameters}
\label{parametervalues}
\begin{center}
\begin{tabular}{c c c c c}
\hline\\
Parameter &GaAs  &  & AlAs&Ga$_{0.7}$Al$_{0.3}$As \\
\hline
$E_g$ & 1519.2&  & 3130 &2002\\
$m_e$     &0.0665 &  &0.124 &0.084\\
$\gamma_1$&6.85&&3.218&\\
$\gamma_2$&2.1&&0.628&\\
$m_{h\parallel H}$ &0.112 &  &0.26&\\
$m_{h\parallel L}$ & 0.210  &  & 0.386& \\
$\mu_{\parallel H}$ &  0.042 &  & & \\
$\mu_{\parallel L}$ &  0.05 &  & & \\
$m_{hzH}$ & 0.38& & 0.51&0.39 \\
$m_{hzL}$ & 0.09 & & 0.22&0.13\\
$R^*_H$&3.64&&13.32&\\
$R^*_L$&4.3&&19.35&\\
$R^*_e$&5.76&&&\\
$a^*_H$&15.78&&7.03&\\
$a^*_L$&13.265&&4.84&\\
$a^*_e$&9.97&&&\\
$\epsilon_b$ & 12.53 &  & 11.16&12.12\\
\hline\\
\end{tabular}
\end{center}
\end{table}

First we determine the electron energy. For the further
calculations we choose the well of GaAs thickness
$L=51~\hbox{nm}$. For the considered GaAs/Ga$_{0.7}$Al$_{0.3}$As
QW we have (Table \ref{parametervalues}) $ m_W = 0.0665~m_0, m_B =
0.0840~m_0$. Using the values for GaAs from table
\ref{parametervalues} and substituting $L=51~\hbox{nm}$ into eq.
(\ref{deftheta}) we obtain $\theta_0 = 21.59$. With this value we
have $\theta=4.972$ from (\ref{stpar_stud_prost}) and the lowest
electron energy
\begin{equation}
E_{e0} = 4R^*_e\left(\frac{\theta a_e^*}{L}\right)^2 =
21.78~\hbox{meV}.
\end{equation}
Quite analogous calculations can be performed for heavy- and light
holes. For the heavy hole one obtains
\begin{equation}
\theta_{0H} = \frac{1}{2}\sqrt{\frac{m_{hzH} V_h}{\mu_{\parl H}
R_H^*}}\frac{L}{a_H^*}.
\end{equation}
Putting the appropriate data from Table \ref{parametervalues} we
have $\theta_{0H} = 20.61$, $\theta_H = 4.99$ and the heavy-hole
energy
\begin{equation}
E_{h0zH} = E_{h0H} = \frac{\mu_{\parl
H}}{m_{hzH}}\left(\frac{2\theta_H a_H^*}{L}\right)^2R^*_H =
4.23~\hbox{meV}.
\end{equation}
For the light hole $\theta_{0L} = 10.82$, $\theta_L=5.27$ and the
energy
\begin{equation}
E_{h0zL} = E_{h0L} = \frac{\mu_{\parl
L}}{m_{hzL}}\left(\frac{2\theta_L a_L^*}{L}\right)^2R^*_L =
17.20~\hbox{meV}.
\end{equation}
Thus the lowest confinement energy for the pair electron-heavy
hole results
\begin{equation}
E_{0zH}=E_{e0z} + E_{h0zH} = 21.78 + 4.23 = 26.01~\hbox{meV}
\end{equation}
and for the pair electron-light hole
\begin{equation}
E_{0zL}=E_{e0z} + E_{h0zL} = 21.78 + 17.20 = 38.98~\hbox{meV}.
\end{equation}
Now we identify the confinement energies with the lowest parabolic
confinement energies:
\begin{equation}
E_{e0} = \frac{\hbar \omega_e}{2}, \qquad E_{h0} = \frac{\hbar
\omega_h}{2},
\end{equation}
and obtain the confinement parameters $\alpha$
\begin{eqnarray}\label{stpar_alphae}
\alpha_ea_H^* &=& \alpha_H^*\sqrt{\frac{m_e\omega_e}{\hbar}}
= \sqrt{\frac{m_e}{\mu_{\parl H}}\frac{E_{e0}}{R_H^*}}=3.07,\\
\label{stpar_alphah} \alpha_ha_H^* &=&
\sqrt{\frac{m_{hzH}}{\mu_{\parl H}}\frac{E_{h0H}}{R_H^*}}=3.08.
\end{eqnarray}
with analogous calculations for the light hole. For the pair
electron-heavy hole (heavy-hole exciton) we obtain
\begin{equation}
v_{0000H} = \frac{V_{0000}}{R_H^*} = 2
\frac{(\tilde{\alpha}_{eH}\tilde{\alpha}_{hH})\tilde{\alpha}_{eH}^2}{\tilde{\alpha}_{eH}^2+\tilde{\alpha}_{hH}^2}\left[\frac{\tilde{\alpha}_{eH}^4}{\tilde{\alpha}_{eH}^2+\tilde{\alpha}_{hH}^2}\left(\frac{\tilde{\alpha}_{eH}^2\tilde{\alpha}_{hH}^2}{\tilde{\alpha}_{eH}^2
+ \tilde{\alpha}_{hH}^2} + \varpi\right)\right]^{-1/2}
\end{equation}
where
\begin{equation}
\tilde{\alpha}_{eH} = a_H^*\alpha_e, \qquad \tilde{\alpha}_{hH} =
a_H^*\alpha_{hH}.
\end{equation}
Making use of eqn. (\ref{stpar_alphae}), (\ref{stpar_alphah}), and
putting $\varpi = 0.1$, we obtain $v_{0000} =v_0= 1.98$. This
value inserted into eq. (\ref{stpar_kwadratowe00}) gives $x =
0.184$ and $\lambda = 0.545$, and from
(\ref{stpar_varepsilonzero}) the lowest heavy-hole exciton energy
$\epsilon_{0H} = -1.128$.
%\section{The choice of electron-hole states}
The lowest absorption peak observed in Ref. \cite{MillerGossard}
corresponds to the energy 1535 meV, and the highest at about 1750
meV. Our calculations give the lowest heavy-hole exciton energy at
\begin{equation}
E_g + E(\hbox{e}0) + E(\hbox{h}0) + \varepsilon_{0H}R^*_H = E_g +
E_{e0} + E_{h0H} + \varepsilon_{0H}R^*_H \approx 1541~\hbox{meV}.
\end{equation}
The resonance at 1750 meV will be obtained for the state $\vert
\hbox{e}4\hbox{h}4\rangle$, i.e.
\begin{equation}
E_g + \left(4 + \frac{1}{2}\right)\hbar\omega_e + \left(4 +
\frac{1}{2}\right)\hbar\omega_h + \varepsilon_{0H}R^*_H = E_g +
9(E_{e0} +
 E_{h0H}) + \varepsilon_{0H}R^*_H \approx 1749~\hbox{meV}.
\end{equation}
The lowest resonance for the light-hole exciton is at energy
\begin{equation}
E_g + E(\hbox{e}0) + E(\hbox{h}0) + \varepsilon_0 = E_g + E_{{e}0}
+ E_{{h}0L} + \varepsilon_{0L}R^*_L \approx 1553~\hbox{meV}.
\end{equation}
whereas for the state $\vert \hbox{e}2\hbox{h}2\rangle$ we have
\begin{equation}
E_g + \left(2 + \frac{1}{2}\right)\hbar\omega_e + \left(2 +
\frac{1}{2}\right)\hbar\omega_h + \varepsilon_{0L}R^*_L = E_g +
5(E_{e0} + E_{h0L}) + \varepsilon_{0L}R^*_L \approx
1709~\hbox{meV}.
\end{equation}
Thus we conclude that the resonances observed in Ref.
\cite{MillerGossard} come from the confinement states labeled by
quantum numbers  0, 1, 2, 3, 4.

As it follows from the relations (\ref{stpar_konstytuwn3}),
 (\ref{stpar_parzystosc}), and
(\ref{stpar_elementymacierzowe}), the nonvanishing elements
$\langle \hbox{e}r\vert \hbox{h}s\rangle$ will be obtained for the
confinement functions of the same parity, it means that either
 $r=2k, s=2m;
k,m = 0,1,2$ or $r=2k+1, s=2m+1$. The same holds for the potential
matrix elements. With regard to this property we choose the
following 13 electron-hole states  with appropriate renumbering
 (both for heavy- and light-hole exciton)
\begin{align}
\vert \hbox{e}0\hbox{h}0\rangle &\rightarrow \vert 1\rangle, \quad
&\vert \hbox{e}1\hbox{h}1\rangle &\rightarrow \vert 2\rangle \nonumber\\
\vert \hbox{e}2\hbox{h}2\rangle &\rightarrow \vert 3\rangle, \quad
&\vert \hbox{e}3\hbox{h}3\rangle &\rightarrow \vert 4\rangle, \nonumber\\
\vert \hbox{e}4\hbox{h}4\rangle &\rightarrow \vert 5\rangle, \quad
&\vert \hbox{e}0\hbox{h}2\rangle &\rightarrow \vert 6\rangle, \nonumber\\
\vert \hbox{e}0\hbox{h}4\rangle &\rightarrow \vert 7\rangle, \quad
&\vert \hbox{e}1\hbox{h}3\rangle &\rightarrow \vert 8\rangle, \\
\vert \hbox{e}2\hbox{h}0\rangle &\rightarrow \vert 9\rangle, \quad
&\vert \hbox{e}2\hbox{h}4\rangle &\rightarrow \vert 10\rangle, \nonumber\\
\vert \hbox{e}3\hbox{h}1\rangle &\rightarrow \vert 11\rangle, \quad
&\vert \hbox{e}4\hbox{h}0\rangle &\rightarrow \vert 12\rangle, \nonumber\\
\vert \hbox{e}4\hbox{h}2\rangle &\rightarrow \vert 13\rangle,
\nonumber
\end{align}
where the notation means, for example
\begin{equation}
\vert \hbox{e}2\hbox{h}0\rangle =
\psi_{e2}(z_e)\psi_{h0}(z_h),\quad\hbox{etc.}
\end{equation}
The same operation is performed for energies for light and heavy hole excitons
\begin{eqnarray}
E_g + E_{er} + E_{hs} - \hbar\omega - {\rm i}{\mit\Gamma}
&\rightarrow& Ejh,\nonumber\\
E_g + E_{er} + E_{ls} - \hbar\omega - {\rm i}{\mit\Gamma}
&\rightarrow& Ejl,\qquad j=1,2,\ldots,13,
\end{eqnarray}
The potential matrix elements
become now a square matrix
\begin{equation}
V_{rsnj} = \langle \hbox{e}r\hbox{h}s\vert
\exp[-v\left(z_e-z_h\right)^2]\hbox{e}n\hbox{h}j\rangle \rightarrow V_{jl}.
\end{equation}
Using this notation we transform the equations
(\ref{stpar_niemianowane}) into a system of linear equations for
the 13 unknown quantities $\mathcal{Y}_j$
\begin{equation}\label{stpar_Yj}
(k_j^2 + \epsilon_0)\mathcal{Y}_j + v_{11}\frac{\lambda}{\lambda +
\varpi}\mathcal{Y}_j - \frac{\lambda}{\lambda +
\varpi}\sum\limits_n v_{jn}\mathcal{Y}_n = b_j
\end{equation}
where
\begin{equation}
b_j = \frac{\Delta_{LT}}{R^*}\langle \hbox{e}r\vert
\hbox{h}s\rangle\psi_0(\rho_0).
\end{equation}
Equation (\ref{stpar_Yj}) can be written in a matrix form
\begin{equation}
A\mathcal{Y} = B,
\end{equation}
where $\mathcal{Y}$, $B$ are vectors
\begin{equation}
\mathcal{Y} = (\mathcal{Y}_1,\ldots,\mathcal{Y}_{13}), \quad\quad
B=(b_1,...,b_{13}),
\end{equation}
and the matrix elements of $A$ are defined as
\begin{eqnarray}
A_{jj} &=& k_j^2 + \epsilon_0 + \frac{\lambda}{\lambda + \varpi}(v_{11} - v_{jj}), \nonumber\\
A_{jn} &=& -\frac{\lambda}{\lambda + \varpi}v_{jn}, \qquad n\neq
j.
\end{eqnarray}

\section{Results for
G\lowercase{a}A\lowercase{s}/G\lowercase{a}$_{1-
\lowercase{x}}$A\lowercase{l}$_{\lowercase{x}}$A\lowercase{s}
parabolic Quantum Well and discussion}\label{secIV} 
\begin{figure}[h]
\begin{center}
\includegraphics[width=0.9\textwidth]{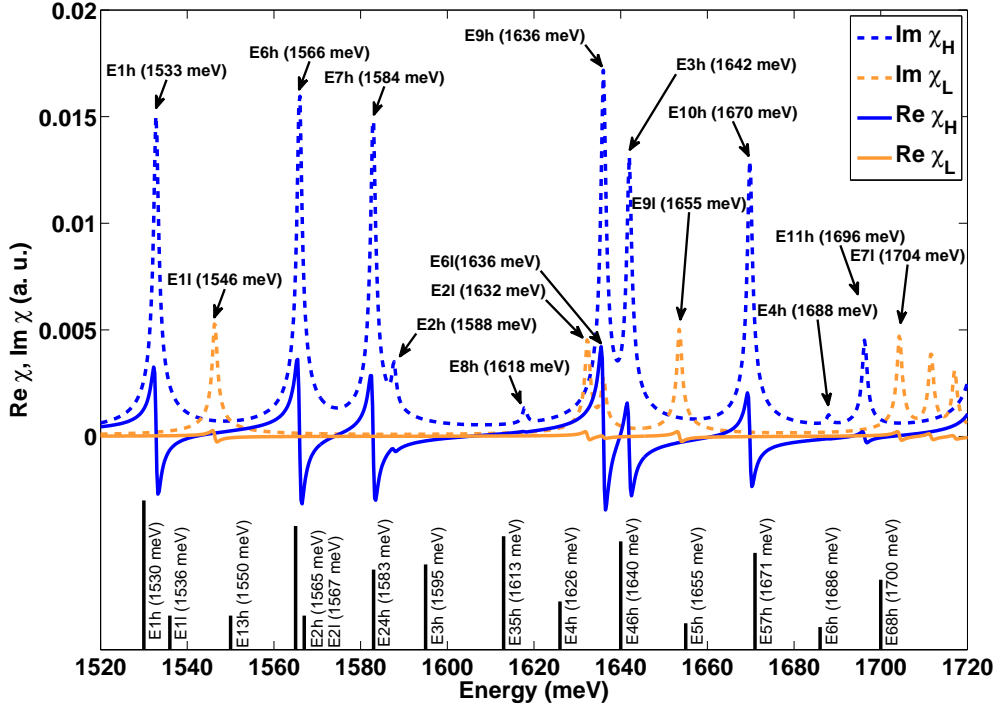}
\caption{The real and imaginary part of the mean QW susceptibility
for the heavy-hole (H) and light-hole (L) exciton. The parameters
used in calculations are $\varpi=0.154,
S=2.6,v=0.5,{\mit\Gamma}=0.5$ and the coherence radii
$\rho_{0L}=0.17~a^*_L, \rho_{0H}=0.1~a^*_H$, respectively. The
electron-hole states and their energies are assessed (indexed by
$1,\ldots, 11$) and the corresponding maxima from
\cite{MillerGossard} are indicated below, with the heights of the
bars indicating the oscillator strengths.}\label{stpar_wynik_fig1}
\end{center}
\end{figure}

\begin{figure}[!ht]
  \begin{minipage}{\textwidth}
    \centering
    \includegraphics[width=.4\textwidth]{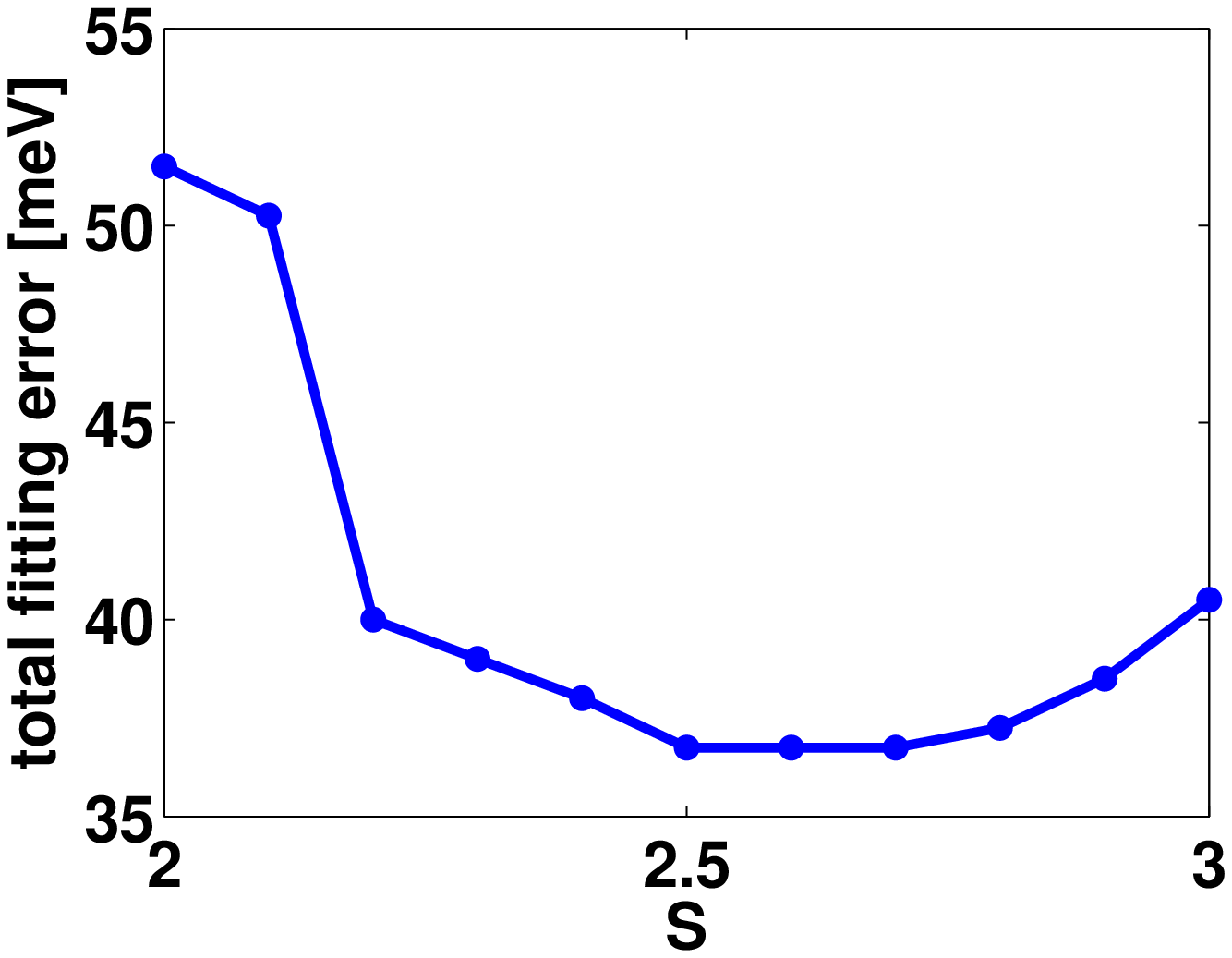}\quad
    \includegraphics[width=.4\textwidth]{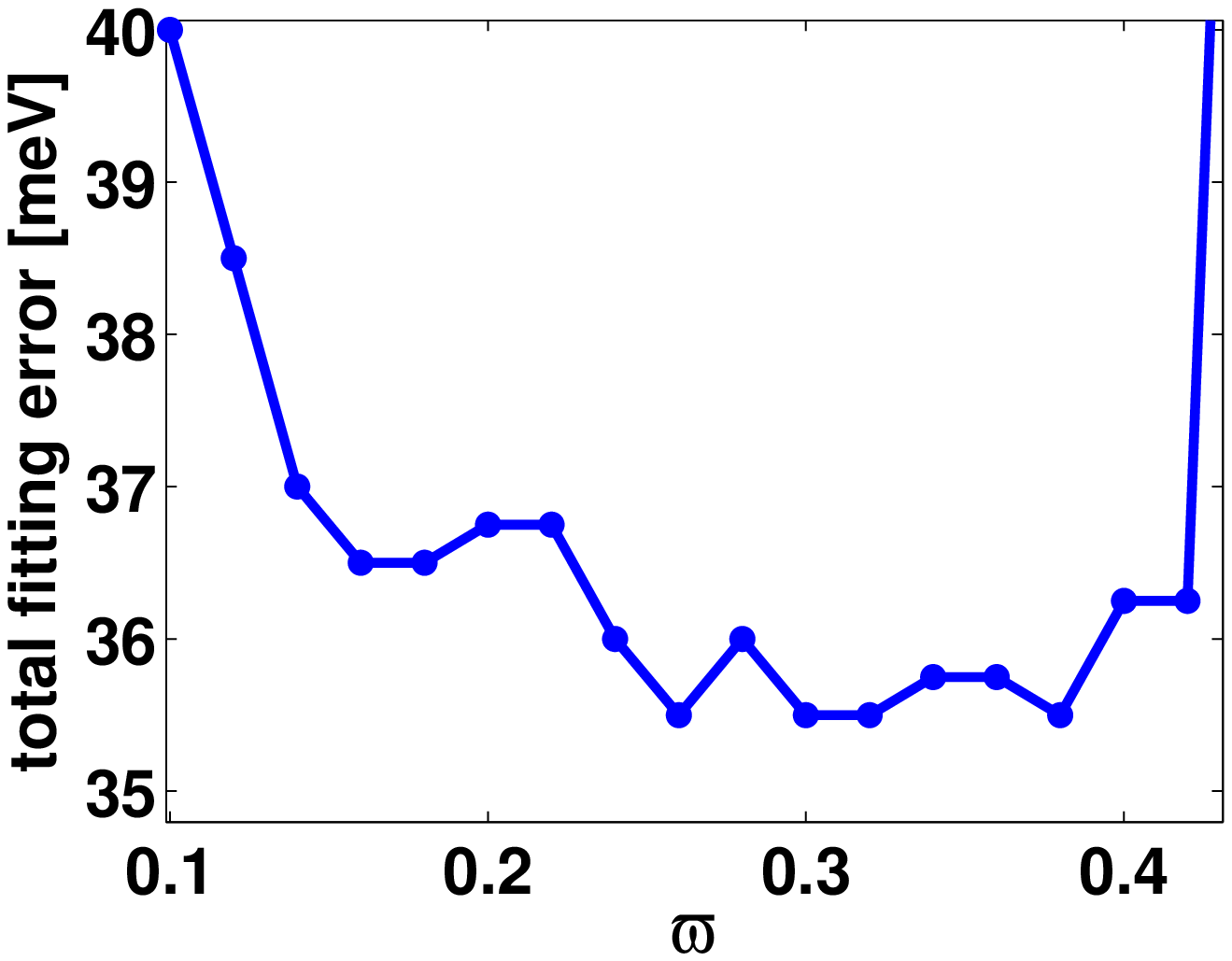}\\
    \subcaption{Total fitting error as a function of $S$ and $\varpi$.}
    \label{fig:sub1}
  \end{minipage}\\[1em]
  \begin{minipage}{\textwidth}
    \centering
    \includegraphics[width=.4\textwidth]{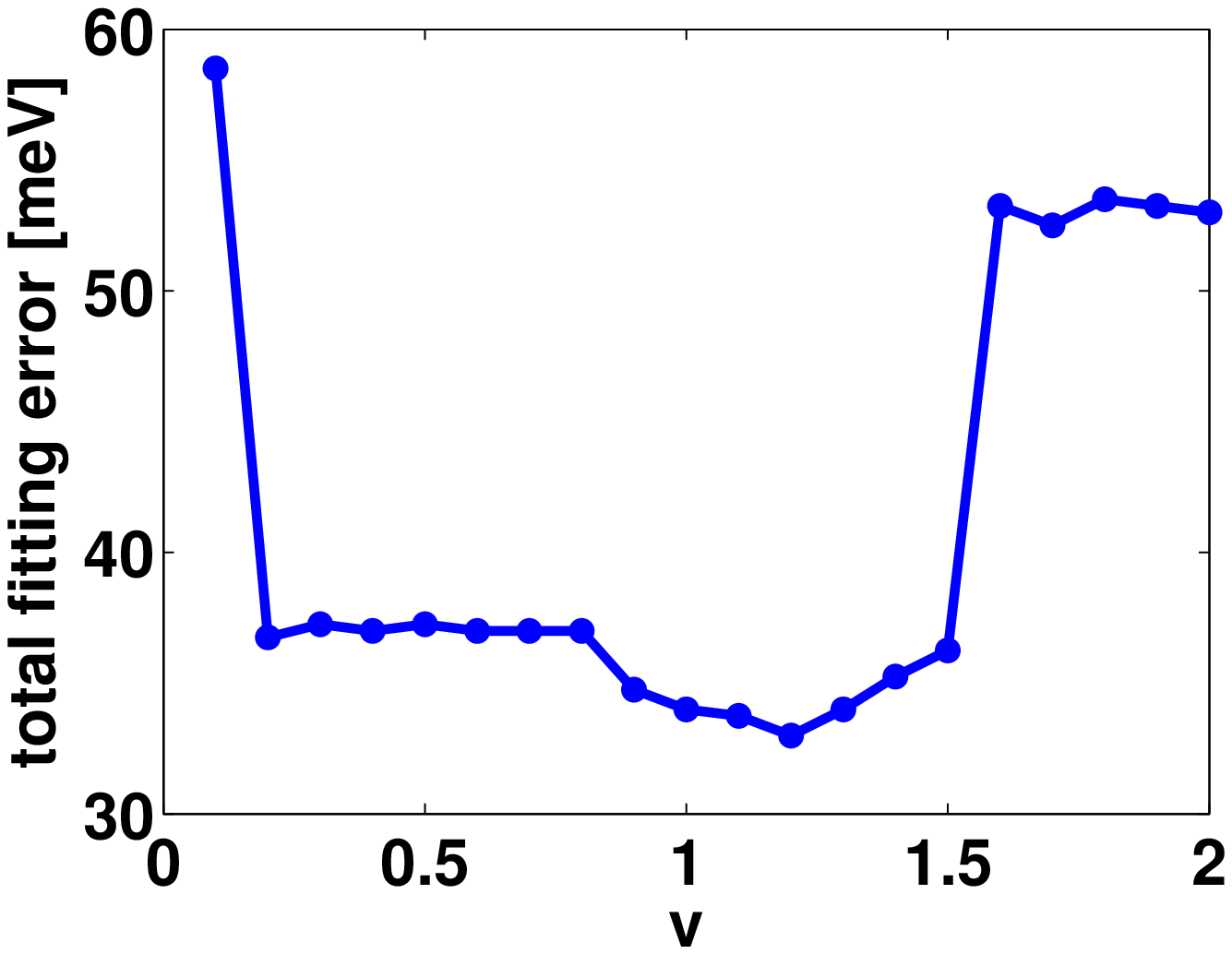}\quad
    \includegraphics[width=.4\textwidth]{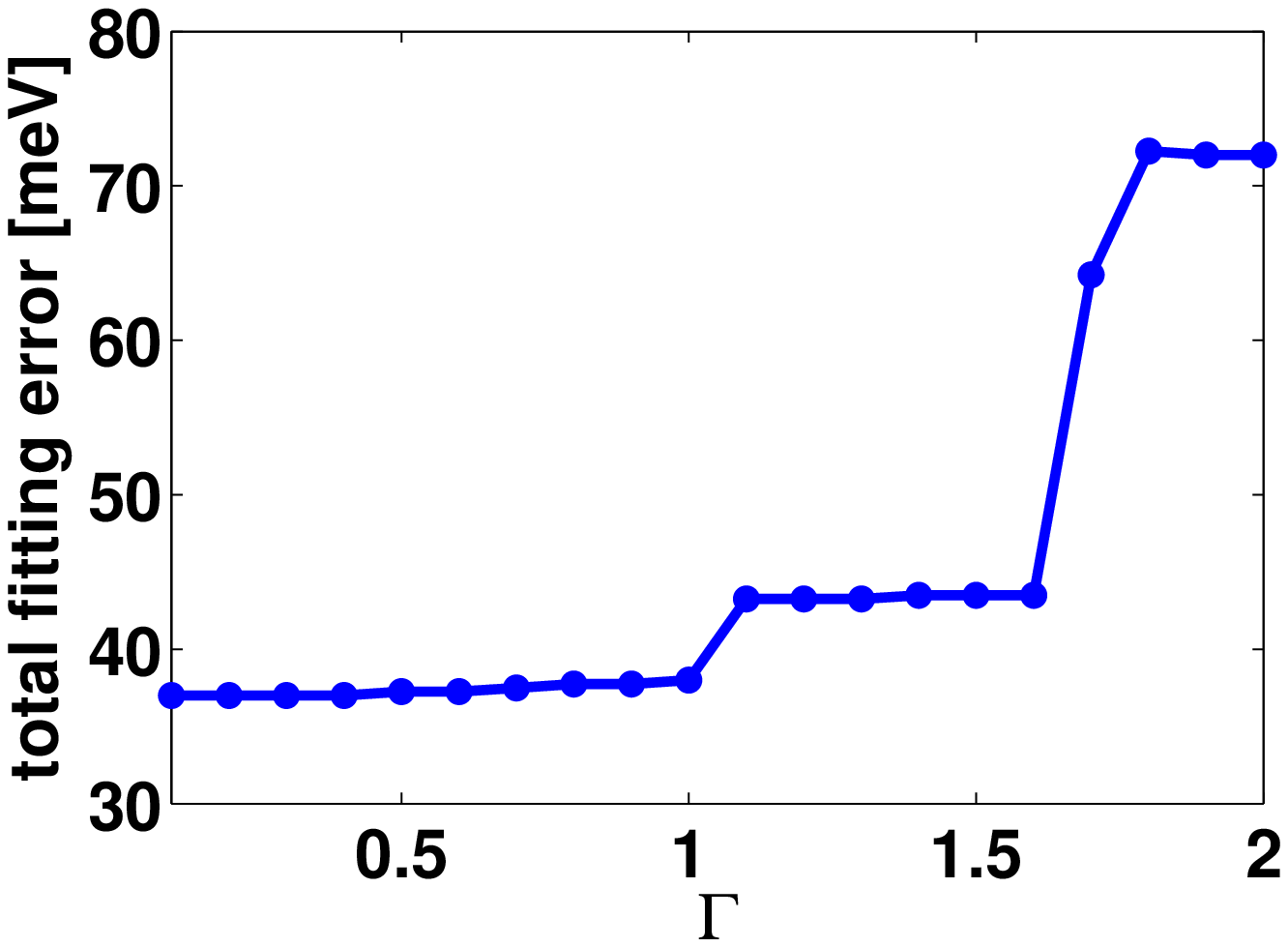}\\
    \subcaption{Total fitting error as a function of $v$ and $\Gamma$.}
    \label{fig:sub2}
  \end{minipage}
    \begin{minipage}{\textwidth}
    \centering
    \includegraphics[width=.4\textwidth]{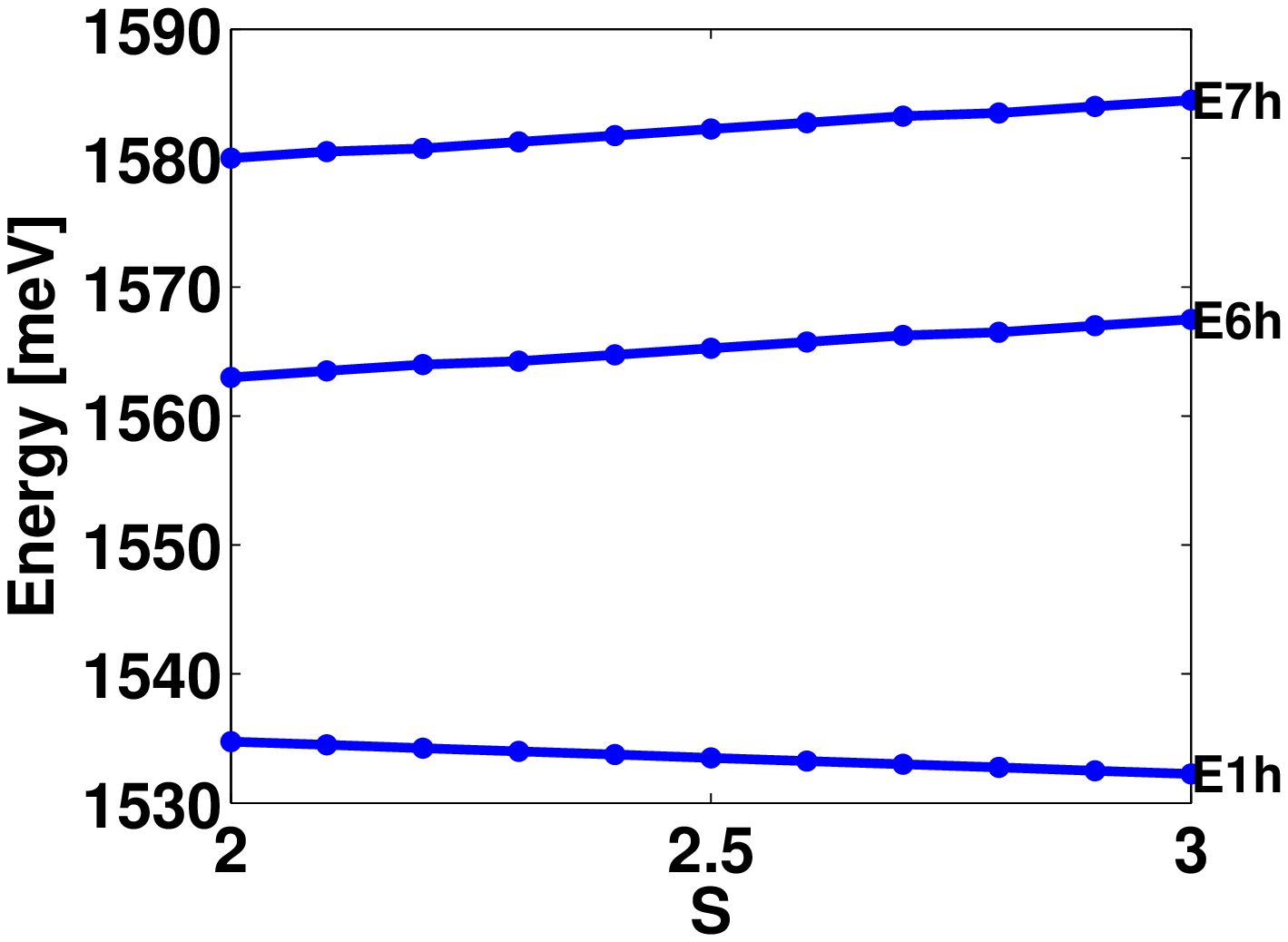}\quad
    \includegraphics[width=.4\textwidth]{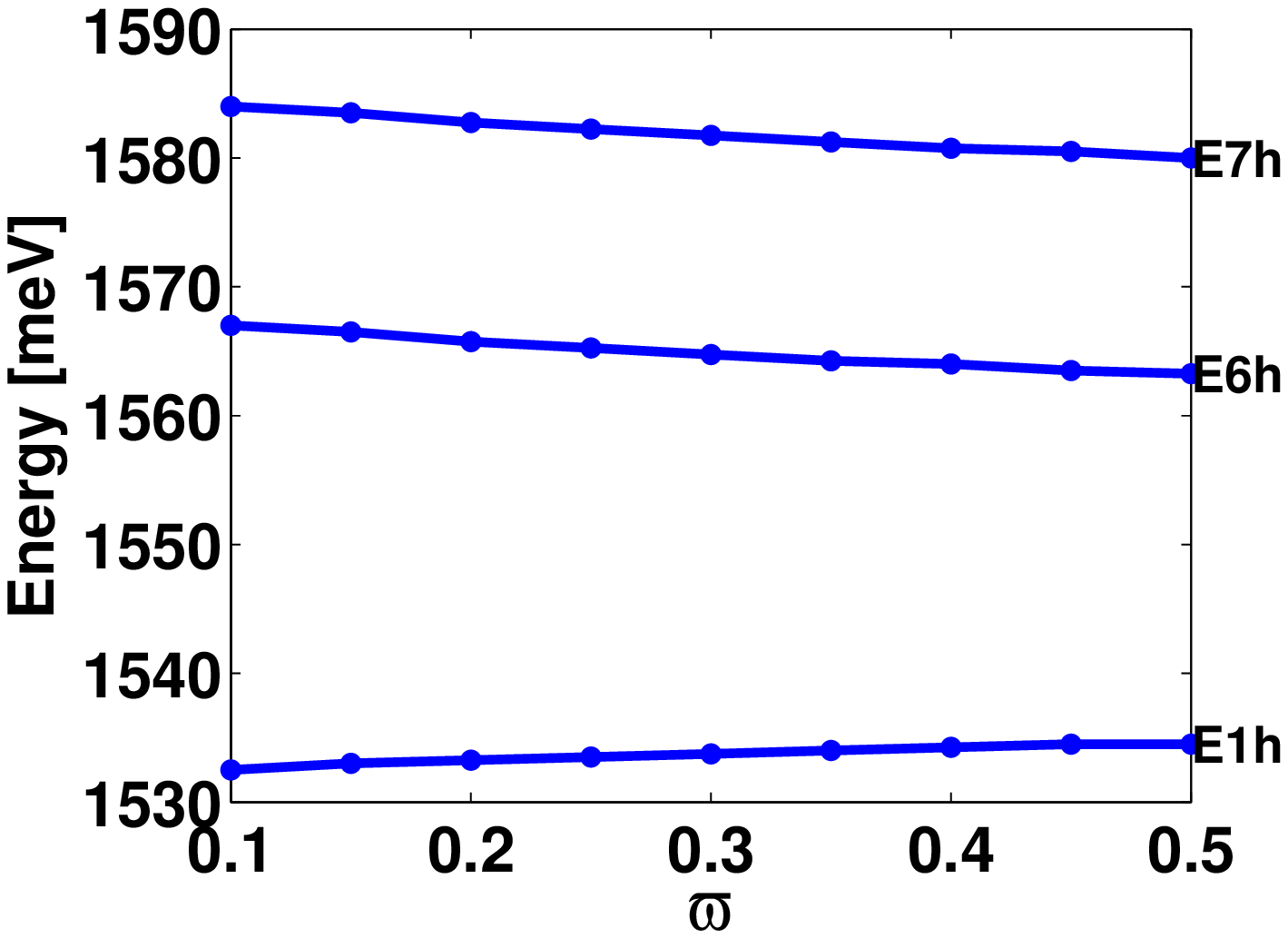}\\
    \subcaption{The effect of the parameters $S$ and $\varpi$ on the position of the first three heavy hole exciton peaks.}
    \label{fig:sub3}
  \end{minipage}
  \caption{The choice of the optimal calculation parameters.}\label{fig:stpar_bledy}
\end{figure}

We have
computed the optical functions of a GaAs/Ga$_{1-x}$Al$_x$As
parabolic Quantum Well with a chosen total thickness of 51 nm. The
values of the relevant parameters are well known, and are given in
Table 1. In our scheme the polarization inside the QW is related
to the coherent amplitudes
\begin{equation}
Y(\rho,z_e,z_h) = \psi_0\sum\limits_{j,n=0}^N\vert
\hbox{e}j\hbox{h}n\rangle Y_{jn},
\end{equation}
by the relation
\begin{equation}
P(z) = 2M_0\psi_0(\rho_0)\sum\limits_{j,n=0}^N\vert
\hbox{e}j\hbox{h}n\rangle(z),
\end{equation}
with the notation
\begin{equation}
\vert \hbox{e}j\hbox{h}n\rangle(z) = \psi_{ej}(z)\psi_{hn}(z).
\end{equation}
Having the polarization, we compute the mean dielectric
susceptibility
\begin{equation}\label{stpar_sr_podatn}
\overline{\chi} =
\pi\epsilon_b\psi_0(\rho_0)\sum\limits_{\ell=0}^N
\mathcal{Y}_\ell\langle 1\vert \ell\rangle_{\Lambda/2}
\end{equation}
where
%\begin{equation}
$\langle 1\vert\ell\rangle_{\Lambda/2} =
\frac{1}{\Lambda}\int\limits_{-\Lambda/2}^{\Lambda/2}\vert\ell\rangle(\zeta){\rm
d}\zeta,  \Lambda=\frac{L}{a^*}$.
%\end{equation}
Having the susceptibility, one can compute, using the appropriate
boundary conditions, the optical functions (reflectivity,
transmission, and absorption). We choose the absorption, which is
related to the effective dielectric function by the formula
\begin{equation}\label{stpar_alpha}
\alpha = \frac{2\omega}{c}\hbox{Im}\sqrt{\epsilon_b +
\overline{\chi}}
\end{equation}
$\epsilon_b$ being the dielectric constant of the QW material. Now
we can compare the theoretical absorption spectra obtained by
(\ref{stpar_alpha}) with the luminescence spectra from
Ref.\cite{MillerGossard}. We have computed the absorption
coefficient for the described above Wide parabolic QW of the
thickness $51~\hbox{nm}$. The first step was to determine the
coefficients $S,\varpi$ satisfying the equations
(\ref{stpar_ukladrownanlamda}). Then, by using the potential
partition (\ref{stpar_podzial_potencjalu}) and the formerly
obtained value  $v$, we have computed the potential matrix elemnts
$V_{rsnj}$ and the matrix elements $\langle r\vert s\rangle$.
Assuming a certain value of the coherence radius $\rho_0$, we have
determined the lowest excitonic eigenfunction $\psi_0$. Finally,
taking a certain value of the damping parameter  $\mit\Gamma$, we
have solved the constitutive equation (\ref{stpar_konstytuwn3}),
obtaining the coherent amplitudes. From the amplitudes we have
computed the mean dielectric susceptibility
(\ref{stpar_sr_podatn}) and the absorption coefficient
(\ref{stpar_alpha}). The results for the real and imaginary part of the mean
susceptibility of the considered QW are displayed in Fig.
\ref{stpar_wynik_fig1}. The parameters used in the calculations
are listed in the figure caption. The arrows indicate the
positions of absorption maxima from Ref. \cite{MillerGossard}. The
good agreement of theory and experiment (both in positions of
maxima and their oscillator strengths) can be seen. In general, we
observe 17 resonance peaks, from which 15 can be identified with
those observed in experiment. The detailed comparison with peaks enumerated by rising energy is shown in the Table \ref{stpar_wynik_tab1}.
%\begin{figure}[htp]
%\begin{center}
%\includegraphics[width=0.7\textwidth]{Fig2.eps} \caption{The real and imaginary part of the mean susceptibility. The %resonances are labeled
% 1$\dots$15, and their positions are listed in Table
%\ref{stpar_wynik_tab1}.  The positions of resonances and the
%corresponding oscillator strengths from ref. \cite{MillerGossard}
%are indicated.}\label{stpar_wynik_fig2}
%\end{center}
%\end{figure}
We have chosen the parameters to obtain the best fit to the experimental results of
Ref. \cite{MillerGossard}. The accuracy of the optimal choice of the effective potential parameters and damping can be tested in the following way. We have computed the total fitting error for the first 13 maxima as a function of the parameters $S$, $\varpi$, $\Gamma$ and $v$. The results are shown on the Fig. \ref{fig:stpar_bledy} (a) and (b). We learned that the positions of the absorption maxima is mainly affected by the values $\varpi$ and $S$. One can see that the change the values of these parameters stretches the whole spectrum, causing a linear shift of the peak position, as shown on the Fig. \ref{fig:stpar_bledy} (c). When using the value $S=2.6$, we obtain $\varpi \approx 0.154$, which represents a local minimum of fitting error. The assumed value of $v=0.5$ is also a good choice. For the global minimum at $v=1.2$, some parts of the absorption spectrum became negative, which was deemed unphysical. As expected, small values of $\Gamma$ have no effect on the location of the peaks. For significant values of $\Gamma$, some peaks become indistinguishable, which is seen as a sudden jump in the fitting error. The selected parameter values gave the theoretical maxima close to the experimental values with mean error of less than 3.5 meV and enabled to identify the electron-hole states.
\begin{table}[h]
\caption{The identification of the electron-hole
states}\label{stpar_wynik_tab1}
\begin{center}
\footnotesize{
\begin{tabular}[htp]{|p{3cm}|p{3cm}|p{5cm}|}
\hline
Number of maximum& State description & Nearest maximum from Ref. \cite{MillerGossard}.\\
\hline
1 & E1h (1533 meV) & E1h (1530 meV), E1l (1536 meV)\\
\hline
2 & E1l (1546 meV) & E13h (1550 meV)\\
\hline
3 & E6h (1566 meV) & E2h (1565 meV), E2l (1567 meV)\\
\hline
4 & E7h (1583 meV) & E24h (1583 meV)\\
\hline
5 & E2h (1588 meV) & E24h (1583 meV), E3h (1595 meV)\\
\hline
6 & E8h (1618 meV) & E35h (1613 meV)\\
\hline
7 & E2l (1632 meV) & E4h (1626 meV), E46h (1640 meV)\\
\hline
8 & E9h (1636 meV) & E46h (1640 meV)\\
\hline
9 & E6l (1636 meV) & E46h (1640 meV)\\
\hline
10 & E3h (1642 meV) & E46h (1640 meV)\\
\hline
11 & E9l (1654 meV) & E5h (1655 meV)\\
\hline
12 & E10h (1671 meV) & E57h (1671 meV)\\
\hline
13 & E4h (1688 meV) & E6h (1686 meV)\\
\hline
14 & E11h (1696 meV) & E6h (1686 meV), E68h (1700 meV)\\
\hline
15 & E7l (1704 meV) & E68h (1700 meV)\\
\hline
\end{tabular}
}
\end{center}
\end{table}

In the next step we tried to fit the experimental line shapes (oscillator
strengths). We have observed that variations of the coherence
radius change substantially the lineshapes. The best fit was
obtained for
 $\rho_{0L}=0.17~a^*_L,\rho_{0H}=0.1~a^*_H$. It can be also verified that the increase of
the damping parameter $\mit\Gamma$ results the lowering of the
oscillator strength.
\clearpage
\section{Conclusions}\label{secV}
We have developed a simple mathematical procedure to calculate the
optical functions of wide parabolic quantum wells. Our procedure
describes  the optical properties of a QW, taking into account the
Coulomb interaction between electrons and holes. Our treatment
includes anisotropic properties of the QW, and takes into account
coherence of the electron-hole pair with the radiation field.
The presented method has been used to investigate the optical
functions of GaAs/Ga$_{1-x}$Al$_x$As parabolic Quantum Well for the
case of radiation incidence parallel to the growth direction and it
shows an excellent agreement with the experimental data,
explaining the number and the positions of the absorption maxima. The justification of the
choice of effective potential parameters and the damping constant is
also presented.
\end{document}